\documentclass[10pt,pra,twocolumn,showpacs,floatfix]{revtex4}

\usepackage{graphicx}
\usepackage{amssymb}
\usepackage{times}
\usepackage{amsmath}
\usepackage{amsthm}
\usepackage{indentfirst}
\usepackage{subfigure}
\usepackage{float}

\begin{document}

\title{Adaptive weak-value amplification with adjustable post-selection}

\author{Fei Li}\author{Jingzheng Huang\footnote{jzhuang1983@sjtu.edu.cn}}\author{Guihua Zeng}
\affiliation
 {State Key Laboratory of Advanced Optical Communication Systems and Networks, Shanghai Key Laboratory on Navigation and Location-based Service, and Center of Quantum Information Sensing and Processing, Shanghai Jiao Tong University, Shanghai 200240, China
}

\begin{abstract}

Weak value amplification (WVA) recently becomes an important technique for parameter estimation, thanks to its ability of enhancing signal-to-noise ratio by amplifying extremely small signals with proper post-selection strategies.
In this work, we propose an adaptive WVA scheme to achieve the highest Fisher information in the case of using an unbalanced pointer. Different from the previous schemes, the adaptive WVA scheme is associated with a real-time update on the post-selection states with the help of the feedback information from outcomes, and the "extremely small" condition set on the parameter of interest is relaxed. By applying this scheme to a time-delay measurement scenario, we show by numerical simulation that the precision achieved in our scheme is several times higher than the standard WVA scheme. Our result opens a new path for improving the WVA technique in a more flexible and robust way.

\end{abstract}

\maketitle

\section{Introduction}

Weak-value amplification (WVA) is a novel metrological technique proposed by Aharonov, Alber and Vaidman\cite{Aharonov1988}.
Since its proposal, several experiments have demonstrated the possibility of WVA scheme to amplify the tiny physical effects \cite{Hosten2008,Gorodetski2012,Dixon2009}.
The experimental successes raise the question of whether weak value amplification can deliver a fundamental advantage for parameter estimation  or whether it should merely be regarded as a convenient experimental tool in certain circumstances \cite{Knee2016}.
The Refs.\cite{Xu2013,Viza2013} have pointed out that the WVA can be used as an advanced technique to provide high sensitivity for parameter estimation.
Furthermore, although an extra bias may be introduced,
it has been proven that the WVA is useful in suppressing certain types of technical noises \cite{Jordan2014,Feizpour2011,Starling2009,Nishizawa2012,Kedem2012}
and in some cases outperforms the standard techniques like interferometric scheme \cite{Brunner2010,Viza2015}.

The WVA scheme generally involves a discrete system and a continuous pointer, which are initialized to states $|\varphi_i\rangle$ and $|\phi\rangle=\int{dx\phi(x)|x\rangle}$ respectively. Here $x$ is a continuous variable, and $|\phi(x)|^2$ is a probability density function with average value of $x_0$ and variance of $\Delta^2$. The system and the pointer are then coupled by an interaction $\hat{U}(g)=\exp[-ig\hat{A}\hat{x}]$ with $\hat{A}$ acting on the system, $\hat{x}$ acting on the pointer and parameter $g$ representing the coupling strength. Finally, before measuring on the pointer, the system is post-selected to state $|\varphi_f\rangle$ that is nearly orthogonal to the initial state $|\varphi_i\rangle$.
After the post-selection, the mean value of the post-selected outcomes in the pointer is shifted by an amount proportional to the weak value given by
\begin{equation}
A_w=\frac{\langle\varphi_f|\hat{A}|\varphi_i\rangle}{\langle\varphi_f|\varphi_i\rangle}.
\label{eq:A_w}
\end{equation}

In the case of using a balanced pointer $x_0=0$ and an ultra small value of $g$ satisfying
$|g| \ll \frac{1}{|A_w|\Delta}$,
there exists an optimal post-selected state $|\varphi^{opt}_f\rangle$ which can concentrate almost all available information about the parameter $g$ into the small fraction of events that survive the post-selection\cite{Alves2015,Zhang2015}.

However, in some other practical scenarios, the pointer is initially unbalanced, such as the detection of spectrum frequency \cite{Brunner2010,LiCF2011}.
In this case, the optimal post-selection strategy is correlated to the value of the parameter of interest.
Moreover, any post-selection strategy deviated from the optimal one will cause a rapid drop-down in estimating precision.
To address the problem, we propose an adaptive weak value amplification (AWVA) scheme which works as follows: in the first run, one roughly measures the value of $g$ with arbitrary pre- and post-selection, and then update the settings by using the outcomes as feedback information. By repeating this process one can achieve the optimal pre- and post-selection for reaching the highest precision. The numerical simulation demonstrates that the precision of our adaptive scheme is improved by several times compared to the standard scheme.

This paper is organized as follows: In Sec.\uppercase\expandafter{\romannumeral2} we study the optimal pre- and post-selection that can obtain all the information contained in WVA scheme.
Based on the result, we further propose a new WVA scheme to adaptively adjust pre- and post-selection to approximately reach the optimum.
In Sec.\uppercase\expandafter{\romannumeral3} we take the time delay measurement for example to illustrate the adaptive WVA scheme and give a numerical simulation result.
A brief conclusion is presented in Sec.\uppercase\expandafter{\romannumeral4}.

\section{Adaptive Weak Value Amplification}

The process of the AWVA scheme is depicted in Fig.1(a) and by comparison the standard WVA (SWVA) scheme is depicted in Fig.1(b).
In AWVA scheme, the pre- and post-selection is adjustable according to an adaptive method, while in SWVA scheme, the pre- and post-selection is fixed and independent with the measurement outcomes.
The implementation of AWVA scheme includes two main steps:
The first step is to perform a WVA measurement with an arbitrary pre- and post-selection. Using the maximum likelihood method, a rough value of the parameter $g$ can be estimated from the detection outcome P(x,g).
The second step is to adjust the pre- and post-selection and then perform the measurement again.
The adjustment is based on an optimization principle, which is dependent on the estimated value of last measurement.
After repeating the two steps for $n$ times, the pre- and post-selection can converge to the optimum and achieve an improved estimation precision.

\begin{figure}[!h]
\centering
\subfigure[]{
\begin{minipage}{8cm}
\centering
\includegraphics[width=0.9\textwidth]{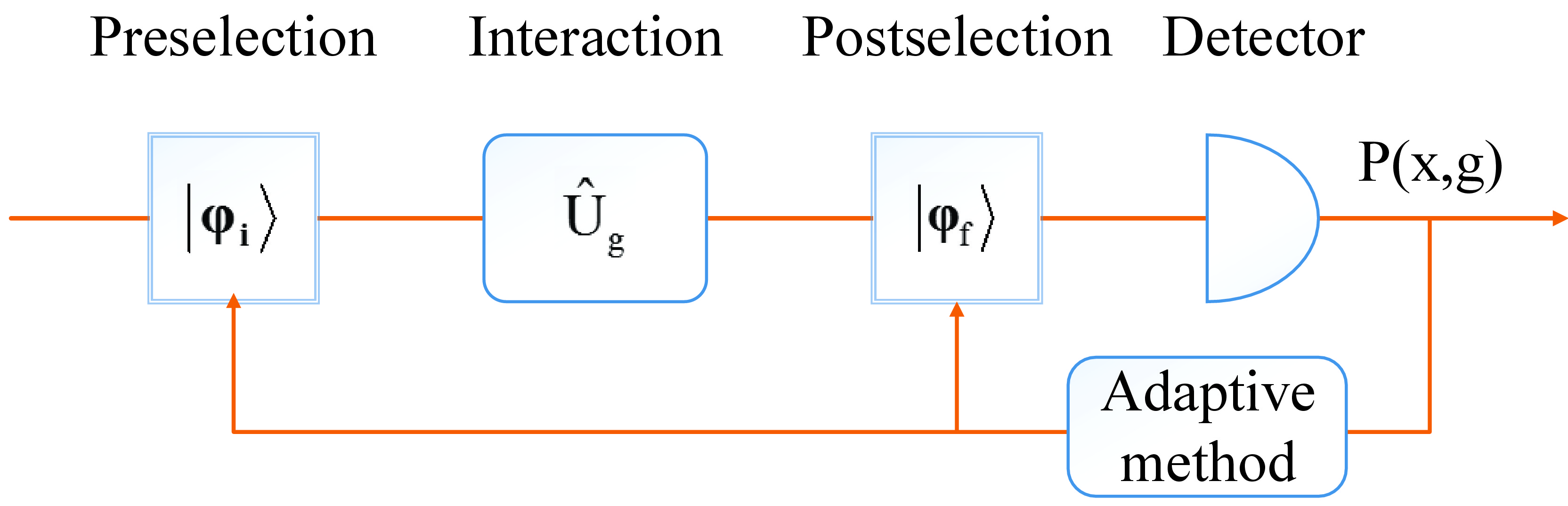}
\end{minipage}
}
\subfigure[]{
\begin{minipage}{8cm}
\centering
\includegraphics[width=0.9\textwidth]{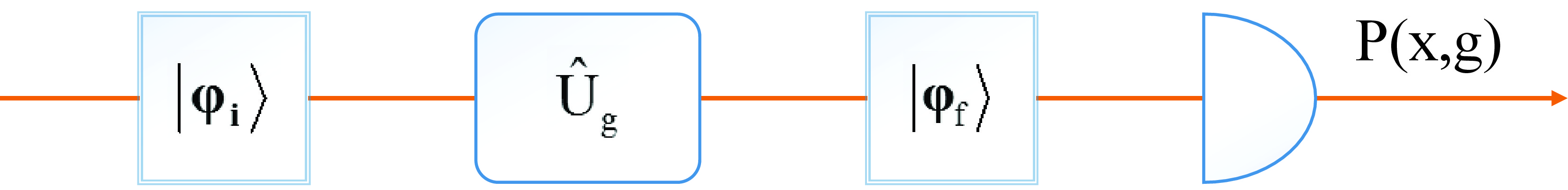}
\end{minipage}
}
\caption{(Color online).
Schematic diagram of the SWVA and the AWVA scheme.
(a) AWVA scheme with feedback control. The pre- and post-selection in each measurement is adjusted according to an adaptive method.
(b) SWVA scheme with fixed pre- and post-selection. The pre- and post-selection is fixed without any adaptivity. The estimative value of coupling parameter $g$ can be calculated from detection outcome.
}
\label{fig:schematic diagram}
\end{figure}

The crucial step in the AWVA scheme is how to adjust the pre- and post-selection in measurement procedure.
By introducing the Fisher information, the problem is converted to choosing the optimal pre- and post-selection that can maximize the Fisher information.
The Fisher information is associated with the important Cramer-Rao bound $Var(g)\geq\frac{1}{\sqrt{NI(g)}}$ \cite{Braunstein1994}, which defines the best attainable precision in estimating the parameter $g$.
Here, $Var(g)=\int{dx(g-\hat{g}_{est})^2P(x,g)}$ is the estimation error averaged over $N$ independent measurement outcomes, $P(x,g)$ is the probability distribution of random variable $x$,
and $I(g)$ is the Fisher information obtained from the measurement and its value is dependent on the employed measurement.
The purpose is to find the best measurement that can maximize the Fisher information in WVA regime and then lead to a lowest limit of estimation precision.

In WVA scheme, as we focus on the data acquired from the successfully post-selected events, the Fisher information $I(g)$ is expressed as \cite{Zhang2015,Fisher}
\begin{equation}
I(g)=P_d\int{dx(\frac{\partial\log{P(x,g)}}{\partial g})^2P(x,g)}
\label{eq:FI}
\end{equation}
It is composed by the information obtained from the measurement on the pointer state after the post-selection and the successful post-selection probability $P_d$ . The normalized probability distribution of $x$ after the measurement on the pointer state is \cite{Fangchen}
\begin{equation}
P(x,g)=|\langle\varphi_f|\varphi_i\rangle|^2P_0(x)\zeta(x,g)/P_d
\label{eq:Probability}
\end{equation}
where $P_0(x)$ is the initial probability distribution of $x$, and $\zeta(x,g)\equiv\cos^2(xg)+\sin^2(xg)|A_w|^2+\sin(2xg)ImA_w$.
The maximum value of $I(g)$ can be obtained by choosing an optimal weak value, as $I(g)$ is a function of $A_w$ through $P(x,g)$ .

\begin{figure*}[!ht]
\centering
\subfigure[]{
\begin{minipage}{8cm}
\centering
\includegraphics[width=1\textwidth]{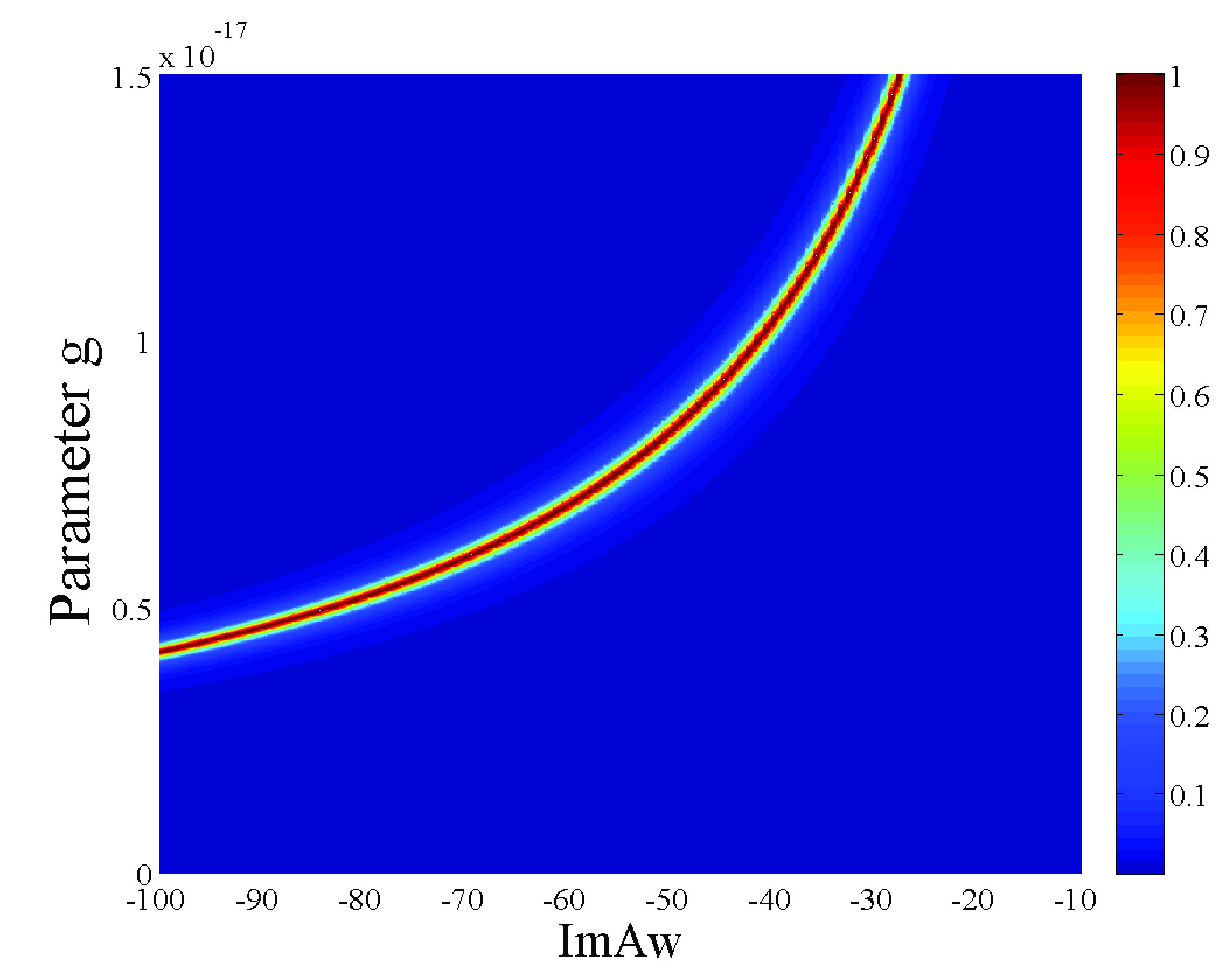}
\end{minipage}
}
\subfigure[]{
\begin{minipage}{7cm}
\centering
\includegraphics[width=1\textwidth]{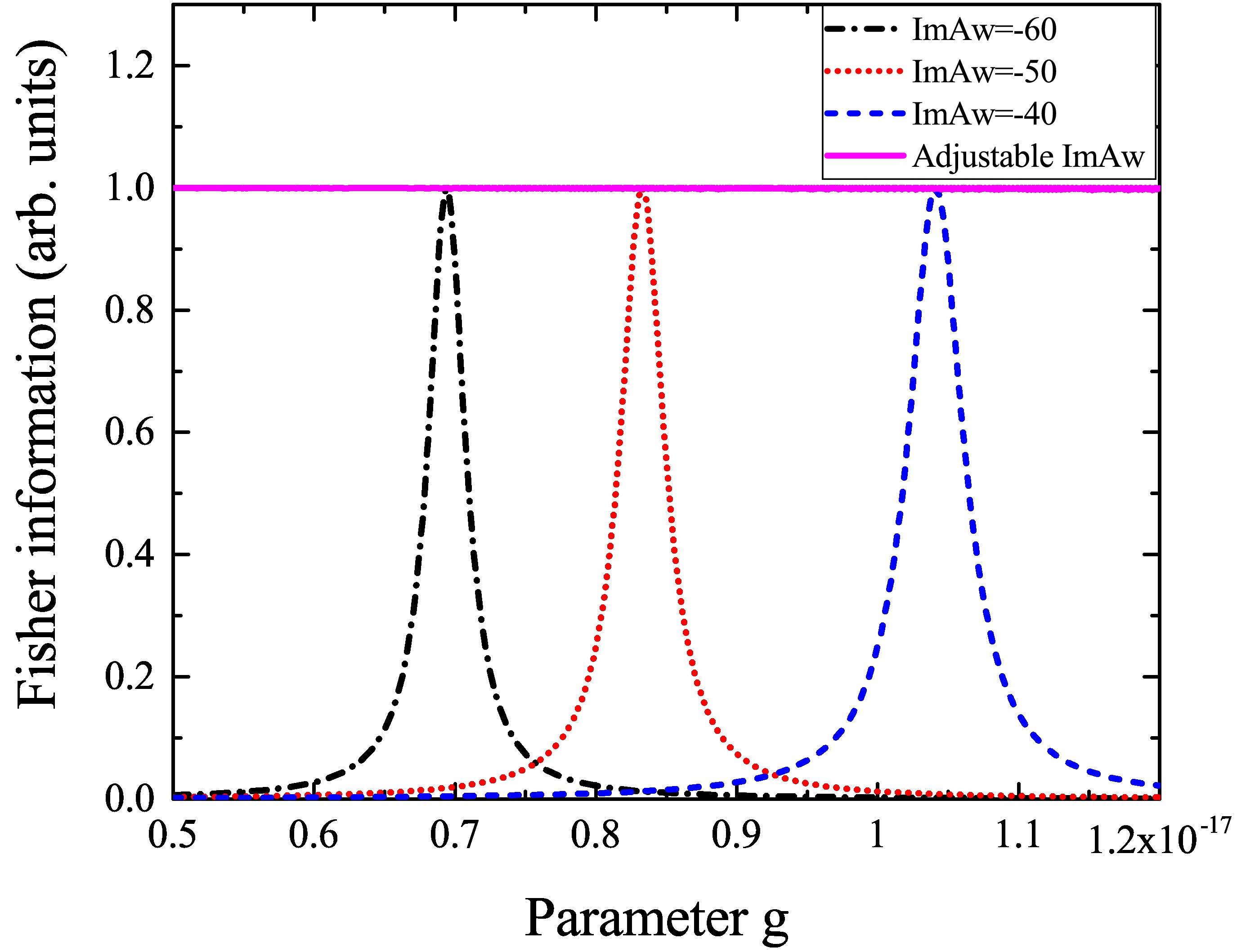}
\end{minipage}
}
\caption{(Color online) (a) The Fisher information (arb. units) with respect to the imaginary part of weak value $ImAw$ and the parameter $g$.
The peak of the curved surface, which denotes the maximum Fisher information, shows that the optimal weak value is varying with the parameter $g$. The figure (b) is captured from this three-dimension surface.
(b) The Fisher information with respect to the parameter $g$. The three dotted curves denote that $ImAw$ takes three different but constant values.
For the three different weak values, denoted by the three dotted curves, the parameter that can achieve the maximum Fisher information is different.
For an alterable weak value, denoted by the solid curve, it can always lead to the maximum Fisher information by adjusting the weak value to be optimal.
}
\label{fig:FI}
\end{figure*}

In a weak-coupling condition $|gx_0|\ll1$, the calculation yields that when the weak value takes a purely imaginary value (the detailed calculation is given in Appendix)
\begin{equation}
\begin{aligned}
\label{eq:condition}
A_w^{opt}=-i\frac{x_0}{\langle x^2\rangle_0 g}
\end{aligned}
\end{equation}
the Fisher information $I(g)$ can reach the maximum
\begin{equation}
\label{I_max}
I_{max}=4\langle x^2\rangle_0
\end{equation}
where $\langle x^2\rangle_0$ is the average of $x^2$.
While with the SWVA condition $|A_wgx_0|\ll1$, the Fisher information is
\begin{equation}
\label{I_wva}
I_{SWVA}=4\Delta^2
\end{equation}
The Eq.(\ref{I_max}) and (\ref{I_wva}) show that the maximum Fisher information
is increased by a factor of $\frac{\langle x^2\rangle_0}{\Delta^2}$ comparing with that in the SWVA scheme.

The maximum achievable information is determined by the quantum Fisher information (QFI), which is valid for all kinds of measurement on the quantum state, including the classical quantum measurement and the weak measurement.
The QFI  corresponding to the joint state  $|\Psi_j\rangle=\hat{U}(g)|\varphi_i\rangle\otimes|\phi\rangle$ is (see Appendix A)
\begin{equation}
Q_j=4\langle x^2\rangle_0
\label{eq:QFI}
\end{equation}
Obviously, the Fisher information $I_{max}$ with optimal $A_w$ can reach the QFI $Q_j$. The result illustrates that if the weak value is chosen appropriately, the full information about $g$ can be obtained by merely considering the post-selected events, in spite of the sharp loss of detection probability.

However, the optimal weak value is associated with the coupling parameter, as is shown by Eq.(\ref{eq:condition}).
This implies that the weak value should vary with the parameter, in order to obtain the maximum Fisher information.
The situation is further interpreted in Fig.\ref{fig:FI}.
(In the graph, we consider the weak value is a purely imaginary number as the real part of weak value plays litter role in our situation.)

In Fig.\ref{fig:FI}(a), the maximum Fisher information is displayed by the peak of the curved surface.
When the Fisher information reaches the maximum, there reveals a clear relevance between the weak value and the parameter.
The Fig.\ref{fig:FI}(b) illustrates it more specifically by capturing three curves from the Fig.\ref{fig:FI}(a).
The three dotted curves represent three different but constant weak values respectively and the solid curve represents a variable weak value.
Clearly, the fixed weak value is the optimum only for a particular parameter value. Once the parameter changes, the Fisher information will decrease sharply. In contrast, the variable weak value, chosen according to Eq.(\ref{eq:condition}), can always achieve the maximum Fisher information.

On the other hand, as the optimal weak value can only be determined if the coupling parameter is already known, the adaptive method, mentioned at the first of this section, is proposed to address the problem.
Based on the optimization condition in Eq.(\ref{eq:condition}),
the pre- and post-selection can finally converge to the optimum through the feedback procedure.

Different from the SWVA scheme constrained by $|A_w g x_0|\ll1$, the requirement of our AWVA scheme is relaxed to $|gx_0|\ll1$.
In this way, the parameter is not limited to be sufficiently small to realize a higher estimation precision.
Besides, comparing with the biased weak measurement scheme which can also realize an ultra-sensitivity measurement for an extremely small parameter $g$ \cite{Zhang2016},
the dynamic range of estimative parameter in our scheme is enlarged by several orders of amplitude .

\section{Time-delay measurement via AWVA}

To specifically illustrate the AWVA scheme, we take the time-delay measurement as an example.
The time delay measurement is to measure a small longitudinal delay $\tau$, which is introduced by a birefringent element between the two pointer states \cite{Brunner2010}. The system is preselected at a fixed state $|\varphi_i\rangle=\frac{|H\rangle+|V\rangle}{\sqrt{2}}$,  where $|H\rangle$ and $|V\rangle$ represent photon polarizations. The pointer is expressed as $|\phi\rangle=\int{d\omega f(\omega)|\omega\rangle}$  with the wave function $f(\omega)=(\pi\sigma^2)^{-1/4}\exp[-(\omega-\omega_0)^2/2\sigma^2]$.
\begin{figure*}[htbp]
\centering
\subfigure[]{
\begin{minipage}{8cm}
\centering
\includegraphics[width=1\textwidth]{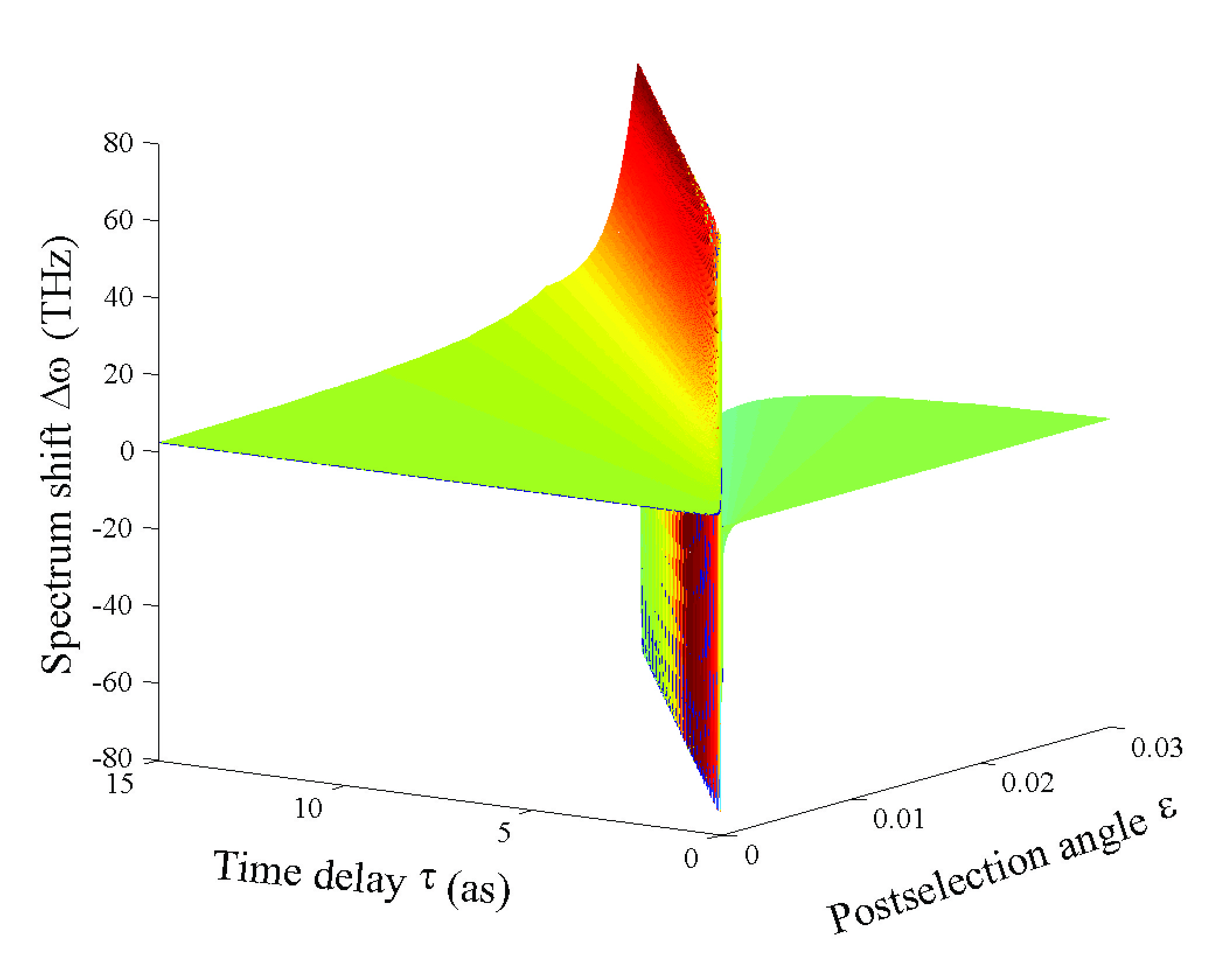}
\end{minipage}
}
\subfigure[]{
\begin{minipage}{6.5cm}
\centering
\includegraphics[width=1.02\textwidth]{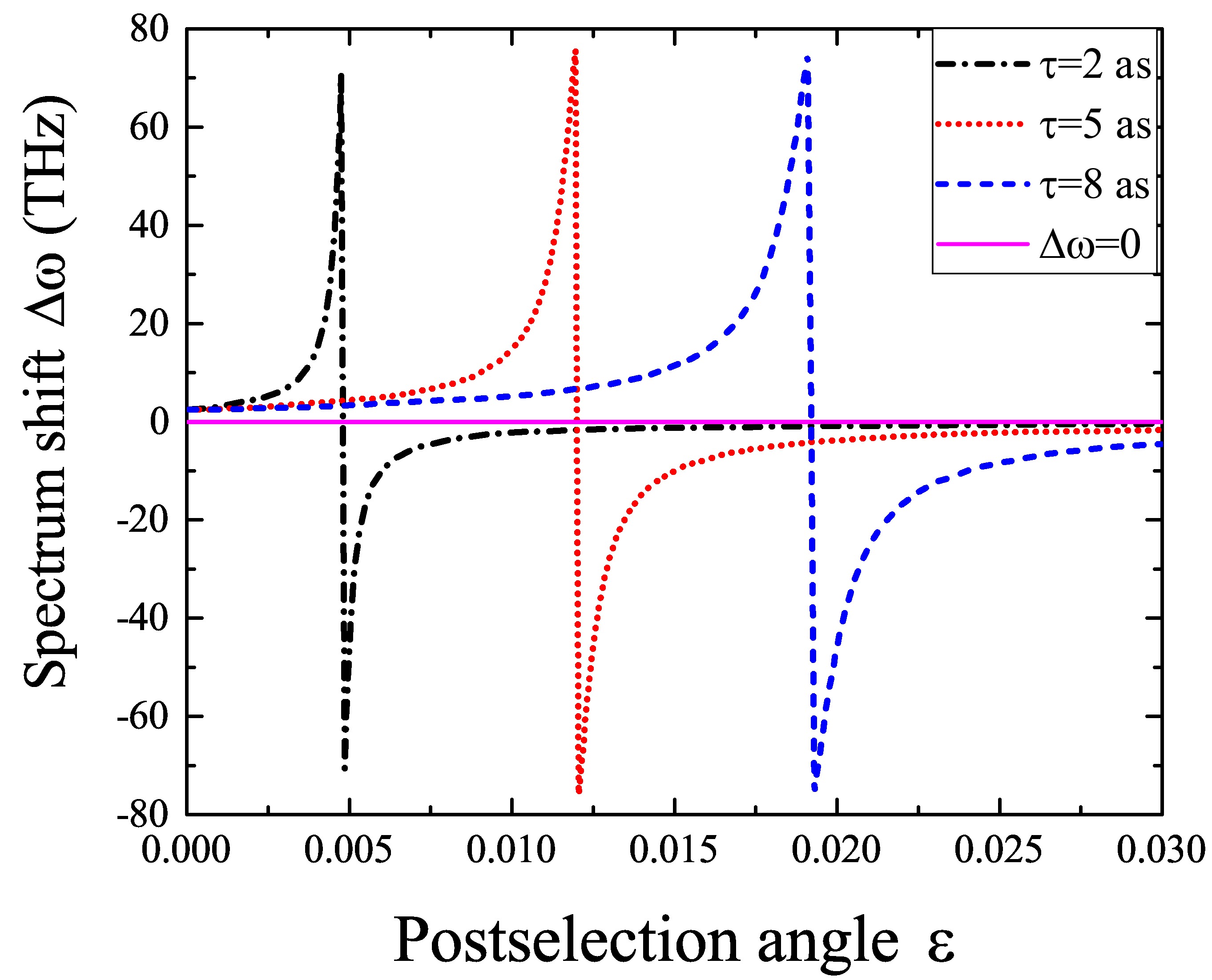}
\end{minipage}
}
\caption{(Color online) (a) Spectrum shift with respect to post-election angle $\epsilon$ and time delay $\tau$. The figure (b) is captured from this curved surface.
(b) Spectrum shift with respect to $\epsilon$.  The three curves denote that when time delay take three different values, the relevance between the spectrum shift and the time delay. Note that the sign of spectrum shift is opposite on different sides of the zero point.}
\label{fig:shift}
\end{figure*}
Undergoing a time-delay process, the system-pointer joint state evolves to
\begin{equation}
\begin{aligned}
|\Psi_j\rangle=\int{d\omega\frac{1}{\sqrt{2}}f(\omega)[e^{-i\omega\tau/2}|H\rangle+e^{i\omega\tau/2}|V\rangle]|\omega\rangle}
\end{aligned}
\end{equation}
Afterwards, a post-selection procedure represented by $|\varphi_f\rangle=\frac{|H\rangle-e^{i\epsilon}|V\rangle}{\sqrt{2}}$ will be performed on the joint state, where $\epsilon$ is an adjustable post-selection angle. In this way, we can optimize the measurement by adjusting the post-selection angle.
After the post-selection, the pointer state becomes
\begin{equation}
\begin{aligned}
|\phi_d\rangle=\frac{1}{\sqrt{P_d}}\int{d\omega(-ie^{-i\epsilon/2})f(\omega)\sin((\omega\tau-\epsilon)/2)}|\omega\rangle
\end{aligned}
\end{equation}
The frequency probability distribution is (without normalization)
\begin{equation}
F(\omega)=\sin^2((\omega\tau-\epsilon)/2)|f(\omega)|^2
\end{equation}
Then the spectrum shift $\Delta\omega$ caused by the time-delay is calculated by the shift of the mean spectrum, which is
\begin{equation}
\begin{aligned}
\label{eq:spectrum-shift}
\Delta\omega&=\frac{\int{ \omega F(\omega)d\omega}}{\int{ F(\omega)d\omega}}-\omega_0=\frac{\sigma^2\tau e^{-\sigma^2\tau^2/2}\sin(\omega_0\tau-\epsilon)}{1-e^{-\sigma^2\tau^2/2}\cos(\omega_0\tau-\epsilon)}
\end{aligned}
\end{equation}
From Eq.(\ref{eq:spectrum-shift}) it is easy to find that around the point $\epsilon=\omega_0\tau$,
$\Delta\omega$ can be simplified as
\begin{equation}
\label{eq:deltaomega}
\Delta\omega\approx\frac{2(\omega_0\tau-\epsilon)}{\tau}
\end{equation}

As the post-selection angle $\epsilon$ is usually a small quantity to realize the amplification of tiny system disturbance, the weak value is calculated as
\begin{equation}
A_w=-i\cot\frac{\epsilon}{2}\approx-i\frac{2}{\epsilon}
\label{eq:A_e}
\end{equation}
Thus, the optimal weak value that obtains the maximum Fisher information can be represented by the post-selection angle, which is
\begin{equation}
\epsilon^{opt}=\frac{\langle\omega^2\rangle_0\tau}{\omega_0}
\label{eq:optimalepsilon}
\end{equation}

We can easily find that the zero point of spectrum shift in Eq.(\ref{eq:spectrum-shift}) is almost consistent with the optimal point in Eq.(\ref{eq:optimalepsilon}) under the condition $\Delta\ll\omega_0$.
It is also pointed out that the sensitivity at the region $\epsilon\approx\omega_0\tau$ can be improved by two orders of magnitude \cite{Zhang2016}. This explains why the region can achieve a better precision from another perspective.
The graph of spectrum shift with respect to $\epsilon$ and $\tau$ is depicted in Fig.\ref{fig:shift}.
It shows that the value of the spectrum shift on different side of the point $\epsilon=\omega_0\tau$ exactly takes the opposite sign, which provides us a simpler way to realize the adaptive method.

\begin{figure}[htbp]\center
	\includegraphics[width=0.45\textwidth]{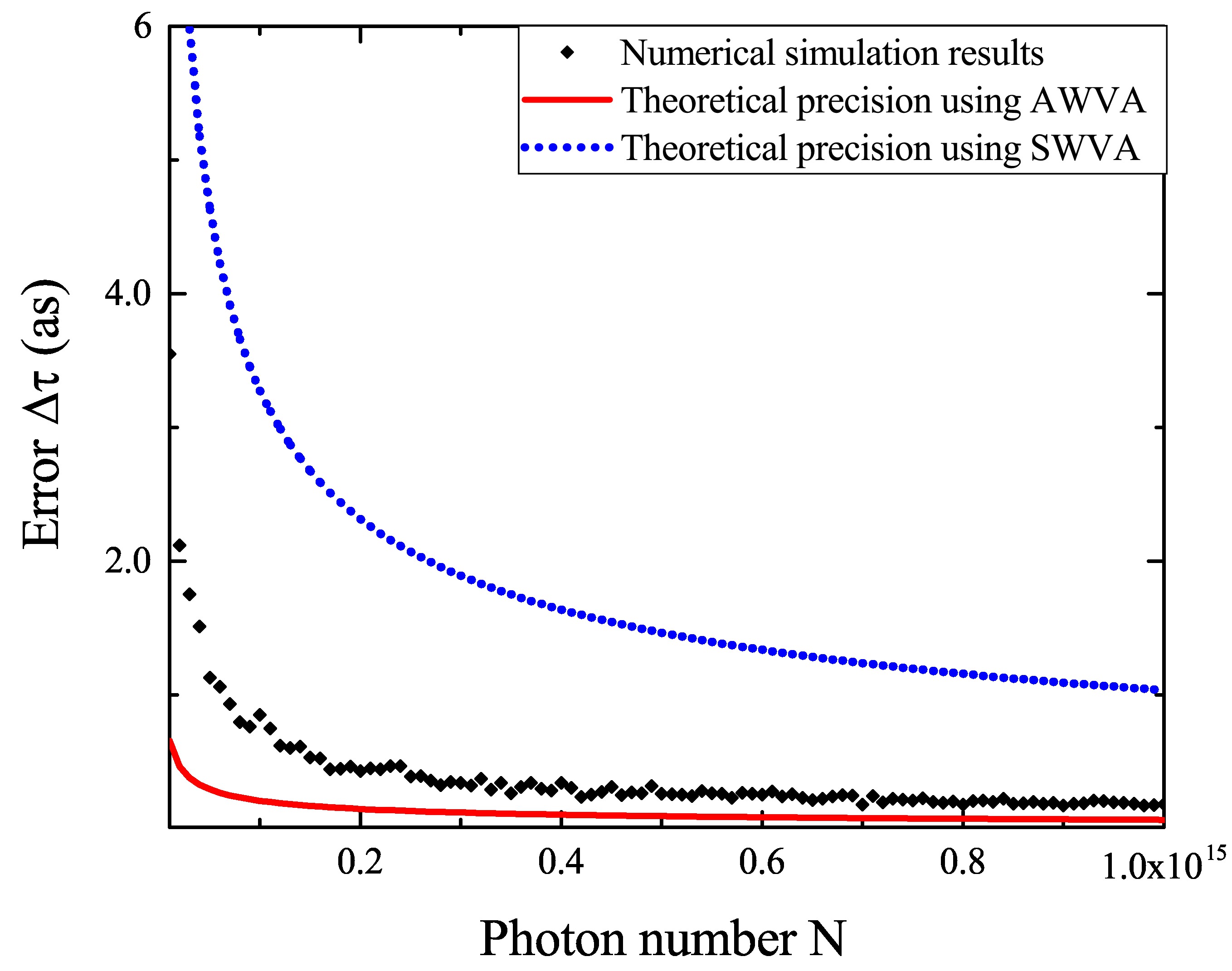}
	\caption{(Color online).
The red continuous curve and the blue dashed curve represent the the minimum errors predicted by the Fisher information of SWVA and AWVA schemes with respect to photon number $N$ respectively.
The black dots are the numerical simulation of the error of AWVA scheme with the optimal post-selection angle for the increasing photon number.
The pre-selection and post-selection are $|\varphi_i\rangle=\frac{|H\rangle+|V\rangle}{\sqrt{2}}$ and $|\varphi_f\rangle=\frac{|H\rangle-e^{i\epsilon}|V\rangle}{\sqrt{2}}$ respectively.
The post-selection angle in SWVA is fixed at $\epsilon = 0.03$ in every run while in AWVA it is nearly adjusted to reach the optimal value via the adaptive method.
The time delay is taken value $\tau_0=8 as$, the mean frequency $\omega_0$ is 2400 THz and the width $\Delta$ is 55 THz. The numerical simulation error is defined as $\Delta\tau=|\hat{\tau}-\tau_0|$  and the error limit is defined as $1/\sqrt{NI}$ . The increment in numerical simulation is $\delta\epsilon=10^{-6}$.
}
	\label{fig:snr}
\end{figure}

In this specific situation, the spectrum shift can serve as the feedback information to adjust the pre- and post-selection.
The measurement procedure is described as following:

1. Perform the WVA measurement with an initial post-selection angle satisfying $\epsilon_0\ll1$, generally within the range $0.001\sim0.1$ \cite{Zhang2016,LiCF2011}, then we will acquire the average spectrum shift $\Delta\omega$ of the pointer.

2. Adjust the post-selection angle according to the $\Delta \omega$.
If $\Delta\omega>0$ , the post-selection angle will be increased by an increment $\delta\epsilon$, i.e., $\epsilon$ will be adjusted to be $\epsilon_1=\epsilon_0+\delta\epsilon$ ;
Otherwise $\epsilon$ will be decreased to be  $\epsilon_1=\epsilon_0-\delta\epsilon$. Then perform the measurement again with the updated post-selection angle $\epsilon_1$.

3. Repeat the above procedures until $\Delta\omega$ changes its sign. Assuming the increment is small enough, the post-selection can be regarded to be the approximately optimal, and the corresponding post-selection angle is $\epsilon^{opt}$. Then the estimated value of time delay $\hat{\tau}$ is calculated by $\hat{\tau}=\frac{2 \epsilon^{opt}}{2\omega_0-\Delta\omega}$ from Eq.(\ref{eq:deltaomega}).

The increment $\delta\epsilon$ is the magnitude of adjustment in every run. The minimum value of it should be smaller than the range of high sensitivity to ensure that the post-selection can be adjusted into the optimal range. It had better be set the minimum allowable value within the experiment setting.

In Fig.\ref{fig:snr}, we numerically simulate the estimation error of AWVA scheme with the photon number increasing.
For reference, we further give the minimum reachable errors determined by the Fisher information in SWVA and AWVA schemes.
As the photon number increases, the simulative error of AWVA scheme will approximately approach the error limit determined by the maximum Fisher information and it keeps less than the error limit of SWVA scheme by several times all the time.
In practice, $\epsilon$ cannot be arbitrarily close to the optimal value because of the finite minimum increment $\delta\epsilon=10^{-6}$, so the simulation error may be higher than the theoretical error limit. Besides, the inherent subtle error in the simulation environment such as the floating error, will also influence the estimation precision. We can optimize the implementation by choosing a smaller post-selection increment $\delta\epsilon$ and increasing the sample size.

\section{Conclusion}

In this paper, we propose an adaptive WVA scheme to achieve the best measurement precision based on the Fisher information theory.
In the weak-coupling region $|gx_0|\ll1$, we derive the optimal pre- and post-selection that can concentrate all the information into the successfully post-selected events with unbalanced pointer.
This optimal pre- and post-selection is related with the parameter of interest.
In this way, the fixed pre- and post-selection in SWVA scheme cannot achieve the maximum information if the parameter changes.
To achieve the maximum Fisher information, we propose a new WVA scheme to adaptively adjust pre- and post-selection via a feedback procedure.
The simulation result shows the estimation precision of AWVA scheme can be improved by several times.
Furthermore, our AWVA scheme can be applied to a broader range of parameter with high precision, comparing with the biased weak measurement scheme in Ref.\cite{Zhang2016}.
In conclusion, our AWVA scheme can be used to estimate a parameter of broader range with a greatly improved precision.

\acknowledgements
We thank to the anonymous referee for their careful work and helpful suggestions.
This work is financially supported by National
Natural Science Foundation of China (Grant No. 61701302).

\appendix
\begin{widetext}
\section{Quantum Fisher information of $|\Psi_j\rangle$}

We first consider the quantum Fisher information (QFI) in WVA measurement with
the system prepared at state $|\varphi_i\rangle=\cos(\theta_i/2)|-1\rangle+\sin(\theta_i/2)e^{i\epsilon_i}|+1\rangle$, and the pointer prepared at state
$|\phi_i\rangle=\int{dpf(x)}|x\rangle$, considering the wave function is a real function. After the interaction expressed by $\hat{U}(g)=e^{-ig\hat{A}\hat{x}}$, the joint state evolves to

\begin{equation}
\begin{aligned}
|\Psi_j\rangle&=\hat{U}|\varphi_i\rangle|\phi_i\rangle\\
&=(e^{-ig\hat{x}}\cos(\theta_i/2)|-1\rangle+e^{ig\hat{x}}\sin(\theta_i/2)e^{i\epsilon_i}|+1\rangle)|\phi_i\rangle\\
&=\int{dx(\cos(\theta_i/2)|-1\rangle e^{-igx}f(x)+\sin(\theta_i/2)e^{i\epsilon_i}|+1\rangle e^{igx}f(x))}|x\rangle
\end{aligned}
\end{equation}

The QFI corresponding to the state $|\Psi_j\rangle$ is generally defined as
\begin{equation}
\begin{aligned}
Q_j&=4[(\frac{\partial\langle\Psi_j|}{\partial g})(\frac{\partial|\Psi_j\rangle}{\partial g})-|\langle\Psi_j|(\frac{\partial|\Psi_j\rangle}{\partial g})|^2]
\label{Q_j}
\end{aligned}
\end{equation}
It stands for all the attainable Fisher information contained in the state $|\Psi_j\rangle$ and it is calculated as following:
\begin{equation}
\begin{aligned}
\frac{\partial |\Psi_j\rangle}{\partial g}=\int{dx(\cos(\theta_i/2)|-1\rangle (-ix)e^{-igx}f(x)+\sin(\theta_i/2)e^{i\epsilon_i}|+1\rangle (ix)e^{igx}f(x))}|x\rangle
\end{aligned}
\end{equation}

\begin{equation}
\begin{aligned}
\frac{\partial \langle\Psi_j|}{\partial g}=\int{dx(\cos(\theta_i/2)\langle -1| (ix)e^{igx}f(x)+\sin(\theta_i/2)e^{-i\epsilon_i}\langle +1| (-ix)e^{-igx}f(x))}\langle x|
\end{aligned}
\end{equation}

\begin{equation}
\begin{aligned}
(\frac{\partial\langle\Psi_j|}{\partial g})(\frac{\partial|\Psi_j\rangle}{\partial g})&=
\int{dx x^2f^2(x)(\cos^2(\theta_i/2)+\sin^2(\theta_i/2))}=\langle x^2\rangle_0
\label{eq:Q1}
\end{aligned}
\end{equation}

\begin{equation}
\begin{aligned}
\langle\Psi_j|\frac{\partial|\Psi_j\rangle}{\partial g}&=\int{dx(\cos^2(\theta_i/2)(-ix)f^2(x)+\sin^2(\theta_i/2)(ix)f^2(x))}\\
&=-i\cos\theta_i\int{ xf^2(x)dx}=-i\cos\theta_ix_0
\end{aligned}
\end{equation}

\begin{equation}
\begin{aligned}
|\langle\Psi_j|\frac{\partial|\Psi_j\rangle}{\partial g}|^2=\cos^2\theta_i(x_0)^2
\label{eq:Q2}
\end{aligned}
\end{equation}
Substitute Eq.(\ref{eq:Q1})(\ref{eq:Q2}) into Eq.(\ref{Q_j}), we finally obtain
\begin{equation}
Q_j=4[\langle x^2\rangle_0-\cos^2\theta_i(x_0)^2]
\end{equation}
Note that when $\theta_i$ satisfies
\begin{equation}
\cos\theta_i=0
\label{eq:theta_i}
\end{equation}
the quantum Fisher information achieves its maximum
\begin{equation}
Q_j=4\langle x^2\rangle_0
\end{equation}

The result shows that the QFI is only related with the initial state of the pointer and system and irrelevant to the specific measurement to be performed.

\section{Fisher information varies with real part of weak value}

To demonstrate the effects on the Fisher information imposed by the real part and the imaginary part of weak value respectively, we define
\begin{equation}
A_w=a+ib
\end{equation}
Firstly, we will devise the correlation between the Fisher information and the real part of $A_w$.
The Fisher information in WVA measurement is defined as
\begin{equation}
\begin{aligned}
I(g)&=P_d\int{dx(\frac{\partial \log{P(x,g)}}{\partial g})^2P(x,g)}
\end{aligned}
\end{equation}
Under the condition $|gx_0|\ll1$ and $|A_w|^2-1\gg1$ as weak value is a large number to realize the signal amplification, $I(g)$ can be simplified as
\begin{equation}
I(g)\approx|\langle \varphi_f|\varphi_i\rangle|^2\int{dx\frac{(\eta'(x,g)\xi(g)-\eta(x,g)\xi'(g))^2}{\eta(x,g)\xi^2(g)}}
\end{equation}
where $\eta(x,g)=1+x^2g^2(a^2+b^2)+2xgb$, $\xi(g)=1+\langle x^2\rangle_0g^2(a^2+b^2)+2\langle x\rangle_0gb$, and $\eta'(x,g)$ and $\xi'(g)$ denote the first-order differential of $\eta(x,g)$ and $\xi(g)$ respectively.

As $|\langle \varphi_f|\varphi_i\rangle|^2$ and $A_w$ are both related with the pre- and post-selection,
we denote $\nu(a)=|\langle \varphi_f|\varphi_i\rangle|^2$ to discuss the problem.
Now we define
\begin{equation}
h(a)=\frac{\nu(a)}{\xi^2(g)}
\end{equation}
\begin{equation}
f(a)=\int{dx\frac{(\eta'(x,g)\xi(g)-\eta(x,g)\xi'(g))^2}{\eta(x,g)}}
\end{equation}
for a clear expression. To obtain the maximum Fisher information, we calculate the differential of $I(a)$ to $a$, that is
\begin{equation}
I'(a)=\frac{\partial I(a)}{\partial a}=h'(a)f(a)+h(a)f'(a)
\end{equation}
where
\begin{equation}
\begin{aligned}
h'(a)=\frac{\partial h(a)}{\partial a}=\frac{\nu'(a)\xi(g)-2\langle x^2\rangle_0g^2a \nu(g)}{\xi^2(g)}
\end{aligned}
\end{equation}
and
\begin{equation}
\begin{aligned}
f'(a)&=\frac{\partial f(a)}{\partial a}\\
&=a\int{dx\frac{(\eta'(x,g)\xi(g)-\eta(x,g)\xi'(g))[2\eta(x,g)Q(a)-2x^2g(\eta'(x,g)\xi(g)-\eta(x,g)\xi'(g))]}{\eta^2(x,g)}}\\
&=aZ(a)
\end{aligned}
\end{equation}
where $Q(a)=4 x^2g\xi(g)+2\langle x^2\rangle_0 g^2\eta'(x,g)-2x^2g^2\xi'(g)-4\langle x^2\rangle_0g\eta(x,g)$, then we have
\begin{equation}
\label{eq:I'}
I'(a)=\frac{\nu'(a)}{\xi(g)}f(a)-af(a)\frac{2\langle x^2\rangle_0g^2 \nu(g)}{\xi^2(g)}+ah(a)Z(a)
\end{equation}

There is not explicit relationship between $\nu$ and $a$, so we define
the preselection state as $|\varphi_i\rangle=\cos(\theta_i/2)|-1\rangle+\sin(\theta_i/2)e^{i\epsilon_i}|+1\rangle$ and the post-selection state as
$|\varphi_f\rangle=\cos(\theta_f/2)|-1\rangle+\sin(\theta_f/2)e^{i\epsilon_f}|+1\rangle$, then we have
\begin{equation}
\begin{aligned}
\nu=|\langle\varphi_f|\varphi_i\rangle|^2&=\cos^2(\theta_i/2)\cos^2(\theta_f/2)+\sin^2(\theta_i/2)\sin^2(\theta_f/2)\\
&=\frac{1}{2}(1+\cos\theta_i\cos\theta_f+\sin\theta_i\sin\theta_f\cos\epsilon_0)
\end{aligned}
\end{equation}
\begin{equation}
\begin{aligned}
A_w&=\frac{\cos(\theta_i/2)\cos(\theta_f/2)-\sin(\theta_i/2)\sin(\theta_f/2)e^{i\epsilon_0}}{\cos(\theta_i/2)\cos(\theta_f/2)+\sin(\theta_i/2)\sin(\theta_f/2)e^{i\epsilon_0}}\\
&=\frac{\cos\theta_i+\cos\theta_f-i\sin\theta_i\sin\theta_f\sin\epsilon_0}{1+\cos\theta_i\cos\theta_f+\sin\theta_i\sin\theta_f\cos\epsilon_0}
\end{aligned}
\end{equation}
where $\epsilon_0=\epsilon_f-\epsilon_i$. Then $a$ and $b$ are expressed as
\begin{equation}
\begin{aligned}
a&=\frac{\cos\theta_i+\cos\theta_f}{1+\cos\theta_i\cos\theta_f+\sin\theta_i\sin\theta_f\cos\epsilon_0}\\
b&=\frac{-\sin(\theta_i)\sin(\theta_f)\sin(\phi_0)}{1+\cos\theta_i\cos\theta_f+\sin\theta_i\sin\theta_f\cos\epsilon_0}\\
\end{aligned}
\end{equation}

With the condition $\cos\theta_i=0$, as discussed in Appendix A, we further get
\begin{equation}
\begin{aligned}
&\nu=|\langle\varphi_f|\varphi_i\rangle|^2=\frac{1}{2}(1+\sin\theta_i\sin\theta_f\cos\epsilon_0)\\
&a=\frac{\cos\theta_f}{1+\sin\theta_i\sin\theta_f\cos\epsilon_0}\\
&b=\frac{-\sin(\theta_i)\sin(\theta_f)\sin(\epsilon_0)}{1+\sin\theta_i\sin\theta_f\cos\epsilon_0}
\end{aligned}
\end{equation}
By calculating the first order differential of $\nu$ to $a$, we get
\begin{equation}
\begin{aligned}
\frac{\partial \nu}{\partial a}&=\frac{\partial\nu}{\partial \theta_f}\frac{\partial \theta_f}{\partial a}\\
&=\frac{-\sin\theta_i\cos\theta_f\cos\epsilon_0(1+\sin\theta_i\sin\theta_f\cos\epsilon_0)^2}{2(\sin\theta_f+\sin\theta_i\cos\epsilon_0)}\\
&=a\frac{-\sin\theta_i\cos\epsilon_0(1+\sin\theta_i\sin\theta_f\cos\epsilon_0)^3}{2(\sin\theta_f+\sin\theta_i\cos\epsilon_0)}\\
&=aK(a)
\end{aligned}
\end{equation}
Then Eq.(\ref{eq:I'}) becomes
\begin{equation}
\label{eq:I1'}
I'(a)=a\frac{K(a)f(a)}{\xi(g)}-af(a)\frac{2\langle x^2\rangle_0g^2 \nu(g)}{\xi^2(g)}+ah(a)Z(a)
\end{equation}
Note from Eq.(\ref{eq:I1'}), we find when $a$ satisfies the condition
\begin{equation}
a=0, I'(a)=0
\end{equation}
This results means that when weak value $A_w$ is a purely imaginary value, the Fisher information can achieve its extreme value.
To satisfy this condition
\begin{equation}
a=\frac{\cos\theta_f}{1+\sin\theta_i\sin\theta_f\cos\epsilon_0}=0
\end{equation}
$\cos\theta_f$ should satisfy
\begin{equation}
\cos\theta_f=0
\label{eq:theta_f}
\end{equation}

Combining the conditions $\cos\theta_i=\cos\theta_f=0$ from Eq.(\ref{eq:theta_i})(\ref{eq:theta_f}), we have $\sin\theta_i\sin\theta_f=\pm1$.
Then $\nu$ and $b$  are simplified as
\begin{equation}
\begin{aligned}
&\nu=\frac{1}{2}(1\pm\cos\epsilon_0)\\
&b=\frac{\mp\sin\epsilon_0}{1\pm\cos\epsilon_0}
\end{aligned}
\end{equation}
By calculation, we get the relationship of $\nu$ and $b$
\begin{equation}
\nu=\frac{1}{b^2+1}\approx\frac{1}{b^2}
\label{eq:nu_b}
\end{equation}
as the weak value is a large value and satisfies $|A_w|\gg1$.

\section{Fisher information varies with imaginary part of weak value}

Considering weak value is a purely imaginary value with $a=0$ to maximize the Fisher information, we define $A_w=ib$.
Combining the condition $|A_w|^2-1\approx|A_w|^2$ and $|\langle\varphi_f|\varphi_i\rangle|^2\approx1/b^2$ in Eq.(\ref{eq:nu_b}), the Fisher information in Eq.(\ref{eq:FI}) can be expanded as
\begin{equation}
\begin{aligned}
I(g)&=P_d\int{dx(\frac{\partial P(x,g)}{\partial g})^2\frac{1}{P(x,g)}}\\
&=|\langle\varphi_f|\varphi_i\rangle|^2\int{dxP_0(x)\frac{[\gamma'(x,g)\xi(g)-\gamma(x,g)\xi'(g)]^2}{\gamma(x,g)\xi^2(g)}}\\
&\approx\frac{1}{b^2\xi^2(g)}\int{dxP_0(g)\frac{[2xb(xgb+1)\xi(g)-(xgb+1)^2\xi'(g)]^2}{(xgb+1)^2}}\\
&=\frac{1}{b^2\xi^2(g)}\int{dxP_0(g)[2xb\xi(g)-(xgb+1)\xi'(g)]^2}\\
&=\frac{4}{\xi^2(g)}\int{dxP_0(g)[x(1+x_0gb)-(x_0+\langle x^2\rangle_0gb)]^2}\\
&=\frac{4}{\xi^2(g)}\int{dxP_0(g)[x^2(1+x_0gb)^2+(x_0+\langle x^2\rangle_0gx)^2-2x(1+x_0gb)(x_0+\langle x^2\rangle_0gx)]}\\
&=\frac{4}{\xi^2(g)}[\langle x^2\rangle_0(1+x_0gb)^2+(x_0+\langle x^2\rangle_0gx)^2-2x_0(1+x_0gb)(x_0+\langle x^2\rangle_0gx)]\\
&=\frac{4}{\xi^2(g)}[\sigma^2(1+\langle x^2\rangle_0g^2b^2+2x_0gb)]\\
&=\frac{4\sigma^2}{1+\langle x^2\rangle_0 b^2g^2+2x_0bg}
\label{eq:I_b}
\end{aligned}
\end{equation}
From Eq.(\ref{eq:I_b}) we find that when $b$ takes value
\begin{equation}
b=-\frac{x_0}{\langle x^2\rangle_0 g}
\end{equation}
the Fisher information $I(g)$ reach the maximum value
\begin{equation}
I_{max}=\langle x^2 \rangle_0
\end{equation}

\end{widetext}

\end{document}